\documentclass[fleqn,10pt]{wlscirep}
\usepackage[utf8]{inputenc}
\usepackage[T1]{fontenc}
\usepackage{amssymb} 
\usepackage{amsmath} 
\usepackage{braket}

\usepackage{hyperref}
\usepackage{cleveref}
\usepackage{color}
\usepackage{float}

\usepackage{pdfpages}
\usepackage[justification=justified]{caption}

\title{Floquet engineering of Kitaev quantum magnets}

\author[1]{Umesh Kumar}
\author[1]{Saikat Banerjee}
\author[1,2]{Shi-Zeng Lin}

\affil[1]{Theoretical Division, T-4, Los Alamos National Laboratory, Los Alamos, New Mexico 87545, USA}
\affil[2]{Theoretical Division, T-4 and CNLS, Los Alamos National Laboratory, Los Alamos, New Mexico 87545, USA}

\begin{abstract}
In recent years, there has been an intense search for materials realizing the Kitaev quantum spin liquid model. A number of edge-shared compounds with strong spin-orbit coupling, such as RuCl$_3$ and iridates, have been proposed to realize this model. Nevertheless, an effective spin Hamiltonian derived from the microscopic model relevant to these compounds generally contains terms that are antagonistic toward the quantum spin liquid. This is consistent with the fact the zero magnetic field ground state of these materials is generally magnetically ordered. It is a pressing issue to identify protocols to drive the system to the limit of the Kitaev quantum spin model. In this work, we propose Floquet engineering of these Kitaev quantum magnets by coupling materials to a circularly polarized laser. We demonstrate that all the magnetic interactions can be tuned in situ by the amplitude and frequency of the laser, hence providing a route to stabilize the Kitaev quantum spin liquid phase.

\end{abstract}

\begin{document}

\flushbottom
\maketitle
\thispagestyle{empty}

\section*{Introduction}

Light-matter interaction not only provides an important way to probe materials’ properties but also offers exciting opportunities to control the physical properties of quantum materials~\cite{oka_floquet_2019,RevModPhys.93.041002,rudner_band_2020}. It has been demonstrated experimentally that light can induce superconductivity~\cite{Fausti_2011, mitrano_possible_2016, PhysRevX.10.011053}, magnetism~\cite{mentink_ultrafast_2015,PhysRevB.96.014406}, topological states~\cite{PhysRevB.79.081406,wang_observation_2013,mciver_light-induced_2020}, and other novel quantum states of matter~\cite{claassen_dynamical_2017}. One particular fruitful protocol is the so-called Floquet engineering by driving materials periodically with a continuous laser. In this Floquet approach, the coupling between light and matter can tune the microscopic parameters, generate new interactions, and even stabilize entirely new states and excitations that do not exist in equilibrium~\cite{PhysRevB.82.235114,PhysRevX.3.031005}. Inspired by the success of light control of quantum states in the past, we investigate the light-driven transition into the quantum spin liquid in strongly correlated magnetic materials.      
 
Quantum spin liquid (QSL) is one class of highly entangled quantum states of matter, which hosts fractionalized excitations~\cite{savary_quantum_2017,broholm_quantum_2020}. The elegant exact solution of the Kitaev spin model unambiguously shows the existence of a quantum spin liquid~\cite{kitaev_anyons_2006}. The presence of an external magnetic field can open a gap and the model supports the Majorana fermions which have the potential for applications in robust topological quantum computation \cite{RevModPhys.80.1083,sarma_majorana_2015}. Kitaev's work has motivated tremendous efforts both theoretically and experimentally to materialize the Kitaev model in quantum magnets~\cite{hermanns_physics_2018,takagi_concept_2019,motome_hunting_2020}. Jackeli and Khaliullin showed that one can realize the Kitaev spin interaction in edge-shared octahedral structure with strong spin-orbit coupling~\cite{PhysRevLett.102.017205}. Several candidate materials, including RuCl$_3$~\cite{banerjee_proximate_2016} and iridates have been identified and have been extensively studied. Encouraging signs of QSL has been reported~\cite{kasahara_majorana_2018}, but unambiguous identification of the QSL remains a challenge. One obstacle from a theoretical perspective is that there exist competing magnetic interactions originated from the different correlation induced exchange processes in materials, which drives the model Hamiltonian away from the ideal Kitaev QSL model~\cite{PhysRevLett.112.077204,rau_spin-orbit_2016}. Numerical study shows that the Kitaev QSL is only stable in a small region of the model parameter space.~\cite{PhysRevLett.112.077204} Therefore, it is highly desirable to design a protocol for continuously in situ tuning the magnetic interactions in materials to drive the system into QSL phases.             

In this work, we propose to use light to tune magnetic interactions in spin-orbit coupled Mott insulators to favor QSL. We consider the multi-orbital strongly spin-orbit coupled Hubbard model, which describes several candidate materials including RuCl$_3$ and iridates. In a simple picture, the coupling of electrons to light has two effects. First, the periodic modulation of the hopping of electrons between different sites effectively normalizes the hopping parameters. Secondly, electrons can absorb or emit photos during the virtual hopping process, and as a consequence, the energy barrier due to Coulomb repulsion also gets modified. Therefore, the strength and even the sign of magnetic interactions can be controlled by light. In addition, an effective Zeeman field can be generated by circularly polarized light through the process called inverse Faraday effect, which provides a new handle to control the quantum states~\cite{Banerjee2021,Sriram2021}. Here, we derive a low energy spin Hamiltonian from the Hubbard model, both using the numerical exact diagonalization (ED) method and the analytical perturbation theory, and explore in detail how the magnetic interactions can be tuned by the frequency and amplitude of the light. 

\section*{Results}
\subsection*{Multi-orbital system} 
\vspace{0.2cm}

\begin{figure}[t]
\centering
\includegraphics[width=0.5\linewidth]{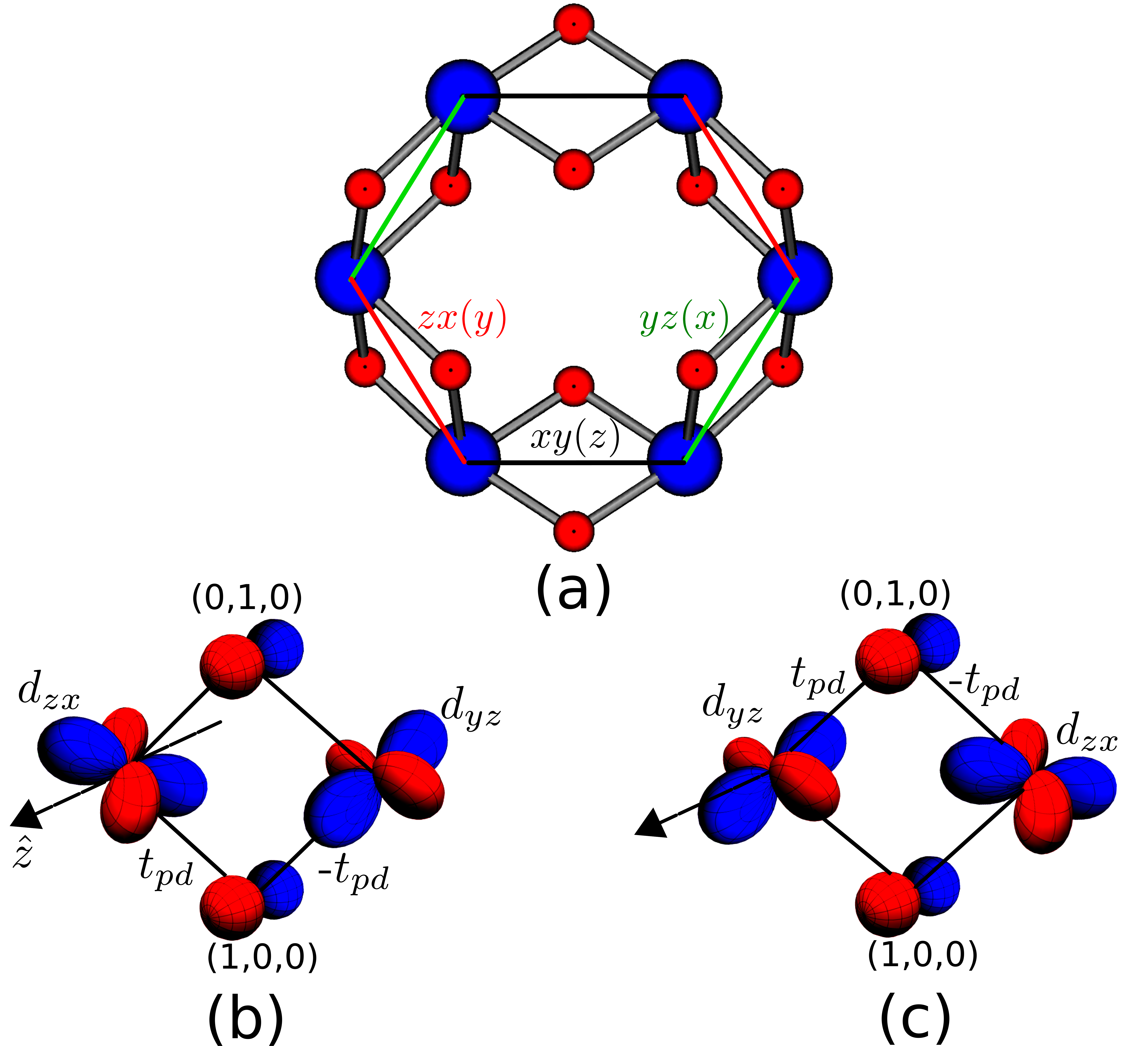}
\caption{Schematics for the edge-shared iridates/ruthenates. Panel (a) shows the full edge-shared hexagon realized in materials. Panel (b) and (c) shows the hopping channels for the $z$-bond mediated by the oxide/chloride [\textit{ligand atoms}] $p$ orbitals. Note that only $d_{yz}$ and $d_{zx}$ orbitals can hop via the ligand $p_z$ orbitals at (1,0,0) and (0,1,0), respectively. Additionally, only the upper and lower hopping paths in panels (b) and (c), respectively, are non-vanishing.}
\label{fig:Schematics_cs}
\end{figure}

We focus on the edge-shared iridates (5d) and ruthenates (4d) as our target materials. In the octahedral crystal field, these materials have been proposed to be described by the $J$-$K$-$\Gamma$ Hamiltonian~\cite{PhysRevLett.112.077204} and as possible candidates to realize Kitaev QSL phase. Here, the $t_{2g}$ manifold of the transition metal (TM) is known to host a hole in the $J_\text{eff}=1/2$ sector, which is separated from $J_\text{eff}=3/2$ states due to a large spin-orbit coupling ($\lambda$) in the system. The $J_{\mathrm{eff}} = 1/2$ multiplet is composed of $d_{xy},~d_{yz}$ and $d_{zx}$ orbitals of the TM atom.  On the other hand, the TM atoms form a two-dimensional (2D) honeycomb lattice perpendicular to [111] direction composed of the standard $x,~y,~z$ - bonds as shown in Fig.~\ref{fig:Schematics_cs}(a). For subsequent discussion, we constrain our analysis for the $z$-bond which is oriented along the [110] direction. The $x$ and $y$ bonds can be recovered by $C_3$-rotation of this bond. Along the $z$-bond, only $p_z$-orbital of the oxygen/chloride ligand is active and have a finite hopping between the $d_{yz}$ and $d_{zx}$ orbitals as shown in panel (b) and (c) of Fig.~\ref{fig:Schematics_cs}, respectively. 
These multi-orbital systems are captured well by Hubbard-Kanamori Hamiltonian and can be mapped to effective spin Hamiltonians using exact diagonalization (ED)~\cite{Mentink2015} or perturbation theory~\cite{PhysRevLett.112.077204, PhysRevLett.121.107201}. The corresponding Hamiltonian is written as
\begin{equation}\label{eq.1}
\mathcal{H}_i^C  =   \frac{U}{2}\sum_{\alpha, \sigma} n_{i\alpha \sigma} n_{i\alpha\bar{\sigma}} +  \frac{U'}{2}\sum_{\substack{\alpha\neq \beta, \\ \sigma, \sigma'}} n_{i,\alpha, \sigma} n_{i\beta \sigma'}  -\frac{J_{\mathrm{H}}}{2}\sum_{\substack{\alpha\neq \beta, \\ \sigma, \sigma'}} d_{i\alpha \sigma}^\dagger  d_{i\alpha\sigma'}^{\phantom{\dagger }}  d_{i\beta, \sigma'}^\dagger  d_{i\beta \sigma}^{\phantom\dagger} + \frac{J_{\mathrm{P}}}{2} \sum_{\substack{\alpha\neq \beta, \\ \sigma}} d_{i\alpha\sigma}^\dagger  d_{i\alpha\bar{\sigma}}^\dagger d_{i\beta\sigma}^{\phantom\dagger}   d_{i\beta\bar{\sigma}}^{\phantom\dagger} + \Delta \sum_{i\sigma} p^{\dagger}_{i\sigma}p_{i\sigma},
\end{equation}
where $U$ and $U'$ are the strength of the intra-orbital and inter-orbital Coulomb repulsion and $J_{\mathrm{H}}$ is the Hund's coupling for the orbitals $\alpha,\beta \in \{ d_{xy}, d_{yz}, d_{zx} \}$, and $J_{\mathrm{P}} = J_{\mathrm{H}}$. $\Delta$ parametrizes the ligand charge-transfer energy. Note that we have ignored the atomic spin-orbit coupling (SOC) term ($\lambda/2 \sum_{i\sigma} d^{\dagger}_{i\sigma} (\mathbf{L}\cdot \mathbf{S}) d_{i\sigma}$), assuming $\lambda \ll U,U',J_{\mathrm{H}},\Delta$. 

Focusing on the four-site cluster [two TM atoms oriented along the $z$-bond with their two edge-shared ligand sites], we now write down the tight-binding Hamiltonian in presence of a circularly polarized light (CPL) as
\begin{equation}\label{eq.2}
\mathcal{H}_K(t)  = \sum_{\langle \alpha, \beta\rangle \sigma} e^{i \delta F (t)} t_{\alpha\beta} d_{1\alpha\sigma}^\dagger d_{2\beta\sigma}^{\phantom\dagger} + t_{pd}\sum_{\sigma} \Big( e^{i\delta { F}_{x} (t)}  d_{1 zx\sigma}^\dagger p_{1\sigma}^{\phantom{\dagger}} - e^{i { F}_{y} (t)}   p_{1\sigma}^{\dagger} d_{2yz\sigma}^{\phantom{\dagger}} 
+  e^{i\delta { F}_{y} (t)} d_{1yz}^\dagger p_{2\sigma}^{\phantom{\dagger}} - e^{i { F}_{x} (t)}   p_{2\sigma}^{\dagger} d_{2zx\sigma}^{\phantom{\dagger}} \Big)+\text{h.c.},     
\end{equation}
where $t_{pd}$ is the hopping amplitude between the $p$- and $d$-orbitals. The hopping matrix between the three $d$-orbitals $t_{\alpha\beta}$ is obtained through Slater-Koster~\cite{PhysRev.94.1498} interatomic matrix elements as
\begin{equation}\label{eq.3}
t_{\alpha\beta} d_{1\alpha\sigma}^\dagger d_{2\beta\sigma}^{\phantom\dagger}  = \begin{bmatrix} d_{1yz\sigma}^\dagger & d_{1zx\sigma}^\dagger & d_{1xy\sigma}^\dagger
\end{bmatrix} \begin{bmatrix} 
t_3 & t_2 & 0 \\
t_2 & t_3 & 0 \\
0 & 0 & t_1 \\
\end{bmatrix} \begin{bmatrix} d_{2yz\sigma} \\ d_{2zx\sigma} \\ d_{2xy\sigma}
\end{bmatrix}.
\end{equation}
The time-dependence of the drive appears as a Peierls phase in these multi-orbital  Hamiltonian~\cite{Mentink2015,UKUMAR2021}. The total Hamiltonian for the two TM-sites is given by $\mathcal{H}(t) = \mathcal{H}_K(t) + \sum_{i =1,2} \mathcal{H}_i^C$. For the sake of simplicity, we consider the light along [001] direction with circular polarization, ${\bf F}(t)= E_0 (\hat{\mathbf{x}} \sin\omega t+ \hat{\mathbf{y}} \cos\omega t)/\omega$. The phase along the TM-TM bond $\left(\tfrac{1}{\sqrt{2}}, \tfrac{1}{\sqrt{2}}, 0 \right)$ is given by $\delta F(t) = \zeta \sin(\omega t- \pi/4)$ and the phases along TM-ligand bonds are given by $\delta F_x(t) = \zeta \sin \omega t /\sqrt{2}$ and $\delta F_y(t) = \zeta \cos\omega t/\sqrt{2}$. Here, the dimensionless parameter $\zeta$ is defined as $\zeta = R_{ij}E_0/\omega$ [$R_{ij}$ is the bond-length between the sites for the associated hopping]. The values of the parameters entering Eq.~(\ref{eq.1})-(\ref{eq.3}) are adapted from the recent \textit{ab-initio}~\cite{PhysRevB.93.155143} and \textit{photoemission}~\cite{Sinn2016} studies relevant for $\alpha$-RuCl$_3$: $t_1 = 0.044$, $t_2 = 0.08$, $t_3 = 0.109$, $t_{pd} = -0.8$, $U= 3.0$, $J_{\mathrm{H}}= 0.45$, $\Delta = 5$. 

Here, we assumed a reduced value of $t_2$ so as to account for finite $t_{pd}$. All the values are in units of eV, unless stated otherwise. We also use relations, $U' = U - 2J_{\mathrm{H}}$ and $J_p = J_{\mathrm{H}}$ in the Hubbard-Kanamori Hamiltonian, which preserves the rotational invariant form. Consequently, the Hubbard-Kanamori term in Eq.~(\ref{eq.1}) can be mapped to 
\begin{equation}\label{eq.4}
\mathcal{H}_i^C = \sum_{i} \bigg[ \frac{U-3J_{\mathrm{H}}}{2} \left( N_i-5 \right)^2 - 2J_{\mathrm{H}} \mathbf{S}_i^2 -\frac{J_{\mathrm{H}}}{2} \mathbf{L}_i^2 \bigg] + \Delta \sum_i n^{p}_{i\sigma},
\end{equation}
where $N_i$ is electron number, $S_i$ is total spin, and $L_i$ is total orbital angular momentum at site $i$~\cite{PhysRevB.95.014409, PhysRevLett.112.077204}. The above simplified form of the Hubbard-Kanamori Hamiltonian [Eq.~(\ref{eq.4})] allows us to evaluate the energy of the doubly occupied states at the TM sites. Since, the SOC $\lambda$ is ignored, we label the doubly occupied sites in terms of the orbital and spin angular momentum $\mathbf{L}$, $\mathbf{S}$, respectively,  as $|^{2S+1}L, L_z, S_z , 0\rangle$, where $L, S~[= (0,0), (1,1), (2,0)]$ are the \textit{orbital} and the \textit{spin} angular momentum of the doublet, respectively, with $L_z$ and $S_z$ being their respective $z$-components. The allowed values of $L$ and $S$ provides three different energies for the doublets (two particles on TM site) given by: $E_S = U+2J_{\mathrm{H}}$, $E_P = U-3J_{\mathrm{H}}$, $E_D = U-J_{\mathrm{H}}$ and singly occupied case (one particle in each TM site), $E_0=0$.

\subsection*{Effective spin-exchange model} 
\vspace{0.2cm}

Since the tight-binding parameters are small compared to interaction strengths [$t_{\alpha\beta},t_{pd} \ll U, U', J_{\mathrm{H}}, \Delta$], the model can be analyzed by a low-energy effective spin-exchange Hamiltonian by ignoring the high-energy charge degrees of freedom. The static limit of our model has been extensively studied previously leading to the well-known $J-K-\Gamma$ spin Hamiltonian~\cite{PhysRevLett.112.077204,Rau_Review2015}. Here, we utilize the Floquet approach to derive a time-independent effective Hamiltonian from Eq.~(\ref{eq.1})-Eq.~(\ref{eq.2}) and subsequently perform ED calculations to estimate the various spin-exchange parameters of the concomitant spin Hamiltonian. On the other hand, the time-independent version of the latter can be naturally captured by a minimal magnetic model [based on the \textit{mirror reflection}, \textit{inversion} and \textit{broken time-reversal} symmetry of TM-ligand-TM atomic cluster associated with the $z$-bond] as
\begin{equation}\label{eq.5}
\mathcal{H}_{\mathrm{eff}} = \begin{bmatrix} S_1^x & S_1^y & S_1^z
\end{bmatrix} \begin{bmatrix} 
J & \Gamma & \Gamma' \\
\Gamma & J & \Gamma' \\
\Gamma' & \Gamma' & J+K \\
\end{bmatrix} 
\begin{bmatrix} S_2^x\\ S_2^y \\S_2^z \end{bmatrix} + \textbf{h} \cdot (\textbf{S}_1 + \textbf{S}_2).
\end{equation}
Since the trigonal distortion is ignored in our model, we have $\Gamma'=0$~\cite{PhysRevLett.112.077204}. The emergent Zeeman magnetic field $\mathbf{h}$ is an effect of the broken time-reversal symmetry due to the applied CPL and is not accounted for in the static version of the associated model. 

We now briefly discuss how the mapping of the multi-orbital model to the spin model is carried out. In the method section, we discuss how the time-dependent Hamiltonian can be mapped to a time-independent form. The time-independent matrix in Bloch structure form is given by, $(\epsilon_\alpha+m\omega)  |\psi_{\alpha, m}\rangle = \sum_{m'} H_{m-m'} |\psi_{\alpha, m'} \rangle $. Here, integer $m$ is the Floquet sector index and $\alpha$ consists of: (i) a single hole in each TM site and an empty ligand site, (ii) two holes [\textit{double occupancy}] on a TM site while the ligand and the other TM site is empty and (iii) one particle each on a ligand and a TM site [see Supplement~\cite{supp} for the exact forms]. The singly occupied states are a product of $J_{\mathrm{eff}}=1/2$ states on the two TM sites, the doublets consists of states $|^{2S+1}L, L_z, S_z , 0\rangle$, where $L, S~[= (0,0), (1,1), (2,0)]$, as mentioned earlier. 

The full Floquet Hamiltonian ($H$) can be written in the eigen-decomposed form as, $H = \sum_\mathfrak{n} E_{\alpha, \mathfrak{n}} |\phi_{\mathfrak{n}}\rangle \langle \phi_{\mathfrak{n}}|$, 
where, $|\phi_{\mathfrak{n}}\rangle  = \sum_{\alpha, m}  a_{\alpha, m}^\mathfrak{n} |\psi_{\alpha, m}\rangle$. Notice that $|\psi_{\alpha', n}\rangle  $  contains all the configuration for the two-site problem. To restrict the basis to $m=0$ and singly occupied basis, one can use a projector; $P_{s, 0}= P_{\alpha\in s} P_{m=0}$. We then have the projected spin Hamiltonian given by; $ H_\text{PG} = P_{s, 0} H P_{s, 0}$. The spin model is valid only when the Floquet band for the singly occupied is well separated from the other Floquet band and the upper Hubbard band. We define $\Delta_l  = \big(E_\text{min}(m=0, s)-E_\text{min-1}\big)/W$ and $\Delta_u  = \big(E_\text{max+1}(m=0, s)-E_\text{max}(m=0, s)\big)/W$ for each $\zeta$ $i.e.~$ $\Delta_u, \Delta_l\ll 1$. Here $E_\text{min}(m=0, s)$ and e $E_\text{max}(m=0, s)$ are the minimum and maximum energy for the singly occupied states in the $m=0$ Floquet sector, and $W = E_\text{max}(m=0, s)- E_\text{min}(m=0, s)$ for a given $\zeta$. 

\begin{figure}[t]
\centering
\includegraphics[width=0.7\linewidth]{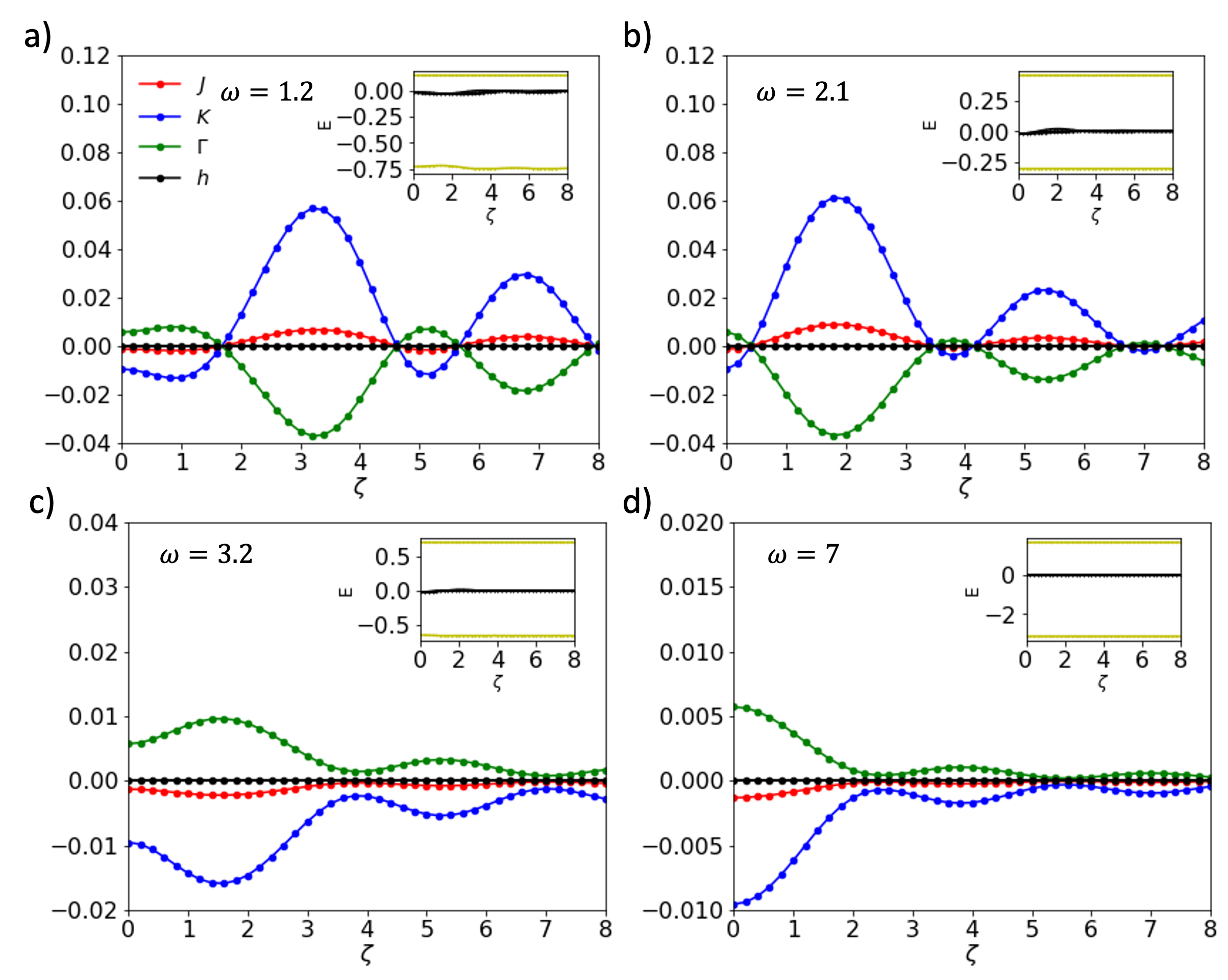}
\caption{Tuning $J$, $K$, $\Gamma$, and $h$ parameters for the model after integrating out the ligand sites. Panels (a) \--- (d) show the result for drive frequency $\omega =1.2, 2.1, 3.2$ and $7$, respectively.  Inset, in each panel, shows the energy levels in the vicinity of $m=0$ Floquet singly occupied states. The black curve (inset) indicates the states with the largest weights on the singly occupied configurations. Note that the vertical axis is plotted in different intervals to highlight the relative strength between different parameters.}
\label{fig:JKGhTuningWLigand}
\end{figure}

Fig.~\ref{fig:JKGhTuningWLigand} and \ref{fig:JKGhTuning_Ligand} plot the estimates using ED for the system without and with ligand, respectively. Inset in the panels of these figures plots the energy states for the $m=0$ Floquet states with singly occupied configuration (in black) and its nearby states. The singly occupied states (in black) are separated from other states, which validates our calculation for the drive frequencies presented in our work. 

\subsubsection*{Spin-exchange model: ED without ligands} 
\vspace{0.2cm}

The ligand degrees of freedom (oxide/chloride $p$ orbitals), in the edge-shared ruthenetes or iridates, are usually integrated out leading to an effective description of the model in terms of the TM $d$-orbitals only~\cite{PhysRevLett.112.077204}. In this section, we perform ED on the Floquet Hamiltonian obtained from $\mathcal{H}(t) = \mathcal{H}_K(t) + \sum_{i=1,2}\mathcal{H}_i^C$ and analyze the results by integrating out the ligand~\cite{supp}. The system without the ligand is simulated by turning off the direct hopping $t_{pd} $ between TM and ligand sites and rescaling the TM-TM hopping $t_{2}$ to $t_2+ t_{pd}^2/\Delta$ in the Eq.~(\ref{eq.2}). Consequently, the four-site cluster effectively becomes a two-site system. In this case, the emergent Zeeman magnetic field $\mathbf{h}$ term is absent as there is no residual orbital current along the TM-TM bond. However, the multi-orbital nature leads to finite non-vanishing spin-exchange couplings $J$, $K$, and $\Gamma$. 

The corresponding results are shown in Fig.~\ref{fig:JKGhTuningWLigand}. We choose four distinct light frequencies avoiding the three critical resonant frequencies $\omega = \{ E_P, E_D, E_S \} [ = \{1.65, 2.45, 3.9 \}]$ of the doublet states and illustrate the relative variation of $J$, $K$ and $\Gamma$ with applied laser strength in the four panels [Fig.~\ref{fig:JKGhTuningWLigand}]. Inset in each panel shows the separation of the $m=0$ Floquet singly occupied states from the adjacent $m\neq 0$ states. Panel (a) shows the variation of the couplings at drive frequency $\omega = 1.2$, below the doublet $E_P$  energy. In this case, we notice that $J, K, \Gamma$ can all change their sign and enhance the magnitude, but the relative tunability is missing. Panel (b) shows the result for $\omega = 2.1$, in between doublet $E_P$ and $E_D$ energies and retains the similar feature as in the prior case. Panel (c) and (d) show the results for $\omega = 3.2$ [$E_D < \omega < E_S$] and $\omega = 7$ [$ \omega > E_S$], respectively. In these two cases, however, we notice that the sign of the parameters cannot be changed, and also the magnitudes of the couplings can be enhanced only mildly. As will be shown by perturbation calculations below, the ratio of $\Gamma/K$ is fixed, independent of the amplitude and frequency of the light.

\subsubsection*{Spin-exchange model: ED with ligands} 
\vspace{0.2cm}

\begin{figure}[t]
\centering
\includegraphics[width=1\linewidth]{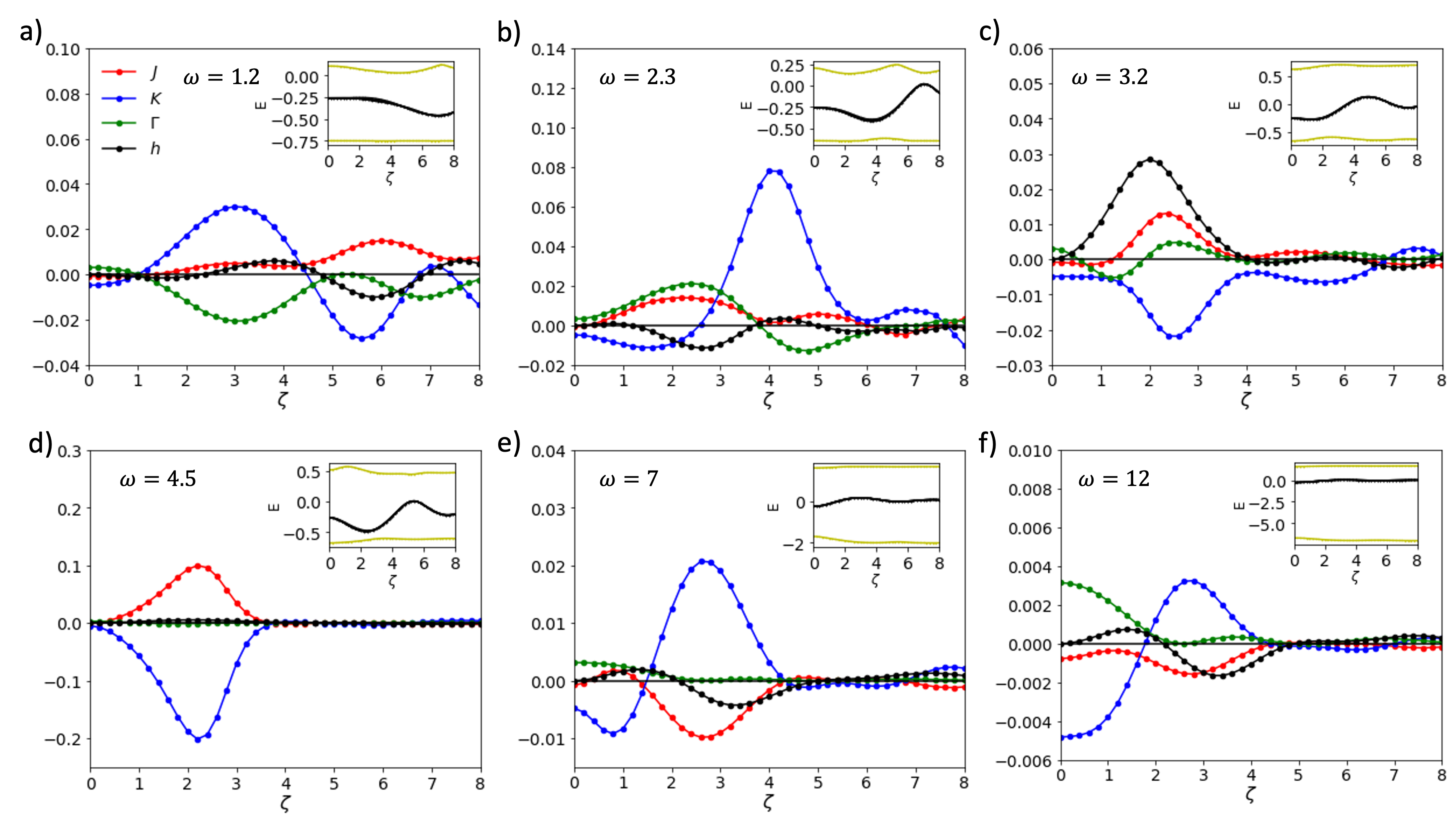}
\caption{Tuning J, K, $\Gamma$ and $h$ parameters in a system with ligands as a function of drive strength ($\zeta$). Panel (a)-(f) show the tuning for drive frequency $\omega = 1.2, 2.3, 3.2, 4.5, 7$ and 12 respectively. Inset in each panel shows the energy levels in the vicinity of $m=0$ Floquet singly occupied states. The black line indicates the states with the largest weights on the singly occupied configurations. Notice that the $y$-axis are plotted in the different intervals to highlight the relative strength between different parameters.}
\label{fig:JKGhTuning_Ligand}
\end{figure}

We now come to the main part of our paper. Here, we present the results using the full Hamiltonian given by Eq.~(\ref{eq.1}) and Eq.~(\ref{eq.2}) for the parameters discussed previously and consider the effect of ligands in full glory. Recently, the inclusion of ligands has been investigated in a few studies~\cite{Chaudhary2020}. The inclusion of ligand degree of freedom in the multi-orbital description is crucial primarily for its two-fold significance. Firstly, in real materials [iridates/ruthenates] the ligand effect is unavoidably present and a thorough quantitative material-specific analysis needs to incorporate it properly. Secondly and most importantly, the inclusion of ligand $p$-orbital along with the TM $d$-orbital offers the desired relative tunability of the spin-exchange couplings - $J$, $K$ and $\Gamma$ [see Fig.~\ref{fig:JKGhTuning_Ligand}], which was absent as we saw in the previous section. Finally, the inclusion of ligand sites constitutes the full TM-ligand-TM atomic cluster with its associated orbital current which leads to a finite Zeeman magnetic field through inverse Faraday effect~\cite{Banerjee2021,Sriram2021}. In Fig.~\ref{fig:JKGhTuning_Ligand} we show the variation of all the parameters defined in Eq.~(\ref{eq.5}) with the driving strength ($\zeta$) evaluated using ED calculations.

As before, we again choose six distinct light frequencies avoiding the four critical resonant frequencies $\omega=\{E_P,E_D,E_S, \Delta \}$ [$= \{1.65, 2.45, 3.9, 5 \}$] of the doublet states and illustrate the relative variation of $J$, $K$ and $\Gamma$ with applied laser strength in the six panels (a)-(f) [Fig.~\ref{fig:JKGhTuning_Ligand}]. The ED analysis is always performed away from the Floquet resonances in order to avoid the breakdown of the spin picture. Inset in each panel shows the separation of the $m=0$ Floquet singly occupied states from the adjacent $m \neq 0$ Floquet states. For field strength $\zeta = 0$ [\textit{static case}], we have $J = 0.0007$, $K = 0.005$, and $\Gamma=0.003$ in each panel, whereas the Zeeman magnetic field $\mathbf{h}$ is zero. The vanishing of the magnetic field is due to the preserved time-reversal symmetry of the multi-orbital system in the absence of CPL.

Panel (a) shows the variation of the various parameters defined in Eq.~(\ref{eq.5}) with the driving amplitude $\zeta$ at the drive frequency $\omega = 1.2$ eV, below the lowest doublet energy $E_P$. Indeed for the reasons mentioned earlier, we find a finite Zeeman magnetic field $\mathbf{h}$ in the driven system, in contrast to vanishing $\mathbf{h}$ in the case without ligand. In this case, the parameters $J$, $K$, $\Gamma$, and $\mathbf{h}$ can all change their sign and their magnitudes can be enhanced. Additionally, these parameters vanish at different values of $\zeta$ [compared to the situation in Fig.~\ref{fig:JKGhTuningWLigand}] allowing for their relative tuning. We also note that the magnitude of the Kitaev term $K$ and anisotropy $\Gamma$ can be enhanced by a few orders of magnitude compared to the Heisenberg exchange term $J$. Panel (b) shows the variation of the parameters with $\zeta$ for $\omega = 2.3$ in between the $E_P$ and $E_D$ doublet. In addition to our findings in panel (a), we notice that $K$ at $\zeta = 3.7$ can be tuned to a few orders of magnitudes larger compared to all the other parameters, making it ideal for realizing the Kitaev QSL phase. The results for $\omega = 3.2$ [$E_D < \omega < E_S$] are shown in panel (c). In this case, the magnetic field $\mathbf{h}$ can achieve the largest magnitude compared to the other parameters. Moreover, we notice limited tunability at larger drive strength $\zeta$. Panel (d) shows the result for $\omega = 4.5$, in the regime between $E_S$ and $\Delta$. In this case, the system can be tuned to large $J$ and $K$, leading to a realization of the Kitaev-Heisenberg model. In addition, the almost vanishing Zeeman field provides a promising route to realize gapless Kitaev QSL. For $\omega =  7$ above the largest energy scale $\Delta$, in our model [see panel (e)], we notice a significant enhancement of $K$ at low drive strength, making it another frequency regime suited for realizing the Kitaev phase.   Panel (f) shows the result for $\omega =  12$, large frequency, well above $\Delta$. In this case, we observe that all the parameters values decrease with the driving strength similar to the case without ligand [see Fig.~\ref{fig:JKGhTuningWLigand}(d)]. In contrast, the tunability of the signs of these parameters persists.

\begin{figure}[t] 
\centering
\includegraphics[width=0.5\linewidth]{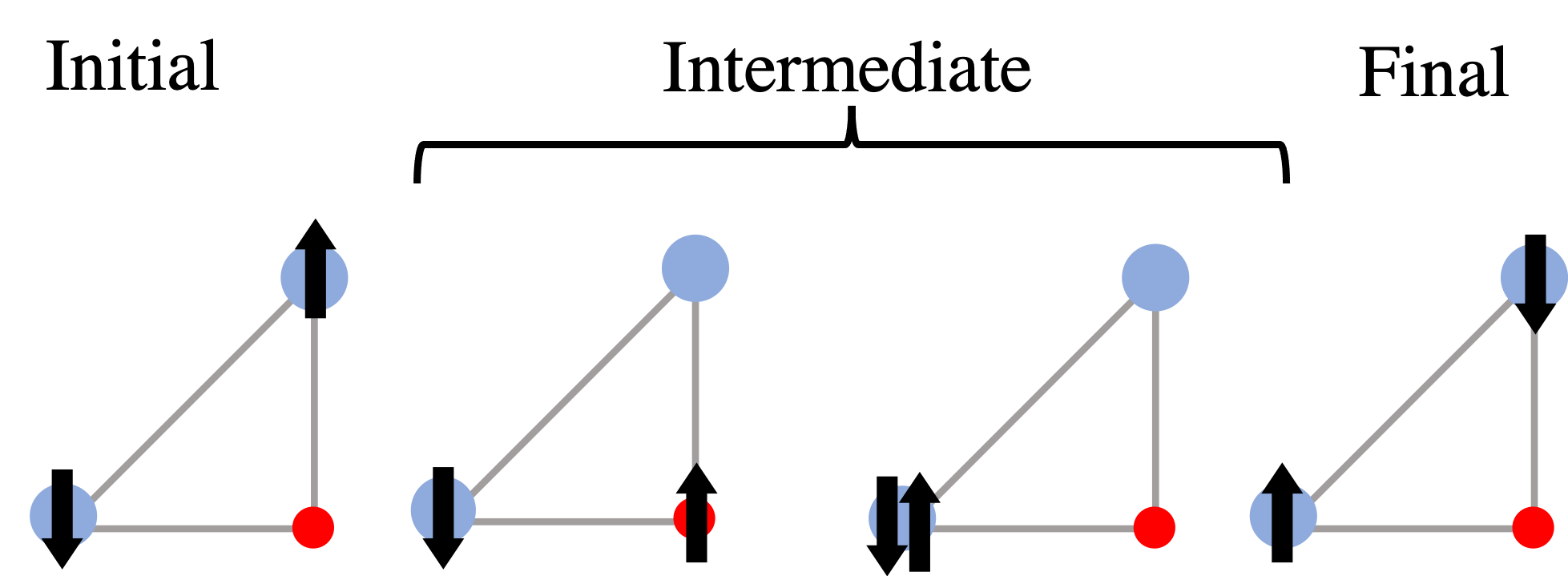}
\caption{Schematics for evaluating spin Hamiltonian parameters at the third-order perturbation theory. We start with one particle each at the TM site. In the intermediates states, the particle moves to a ligand and then to the other TM site creating a doublet. Finally, it returns to the original site, with or without a spin-flip. We also consider a time-reversal partner of this hopping path.}
\label{fig:SchematicThirdOrder}
\end{figure}

\begin{figure}[t] 
\centering
\includegraphics[width=0.7\linewidth]{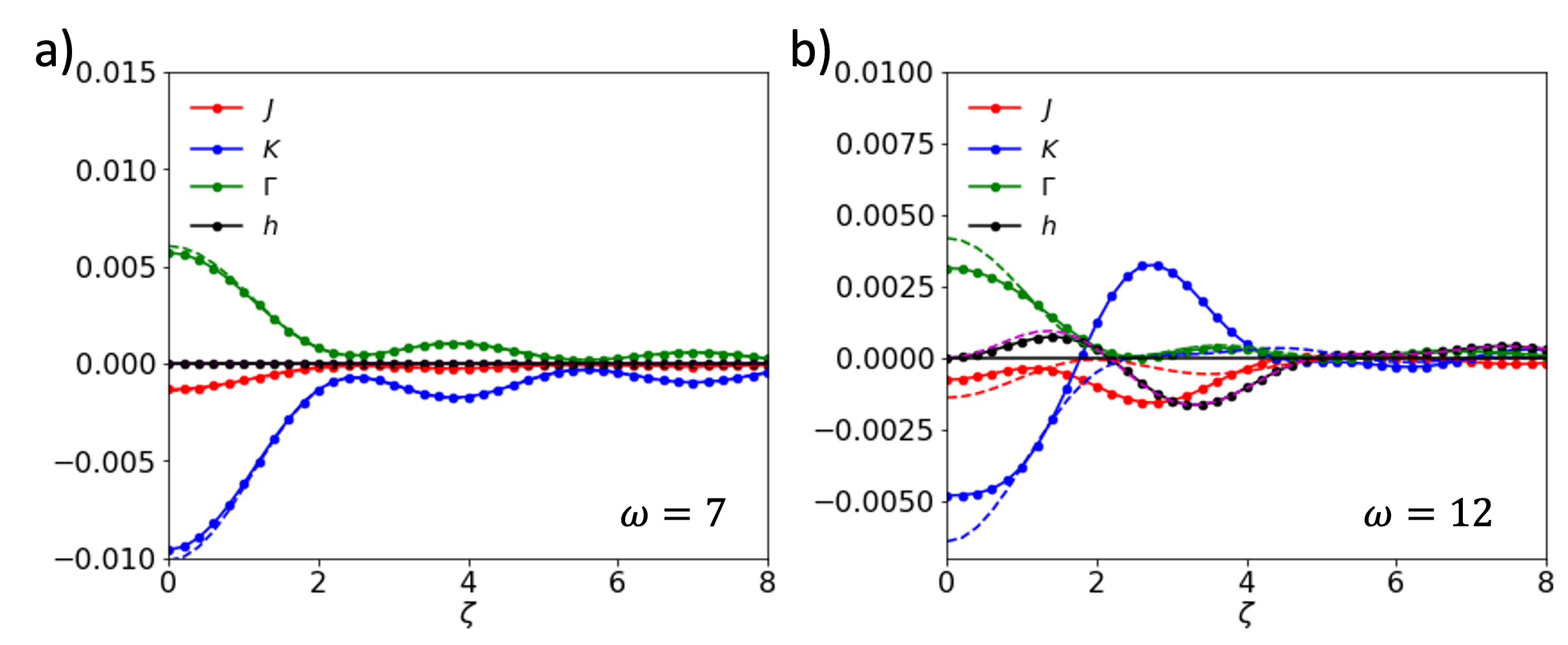}
\caption{Comparison of the ED and perturbation theory results. Panel (a) shows the result for the case without ligand at drive frequency $\omega = 7$. Panel b) shows the comparison in the model with ligands at $\omega =12$.}
\label{fig:EDPertComp}
\end{figure}

\subsection*{Spin-exchange model: Perturbation theory} 
\vspace{0.2cm}

So far we analyzed our model in the Floquet approximation utilizing the ED calculations and discussed the results in various regimes of the model-based parameter values. In this section, we derive the low-energy spin-exchange Hamiltonian based on perturbation theory and compare the analytical expressions of the couplings $J$, $K$, $\Gamma$, and $\mathbf{h}$ with the numerical estimates from the ED. 

In the absence of ligand degree of freedom, we perform a second-order perturbation theory~\cite{supp} and time-dependent Schrieffer-Wolff transformation~\cite{Banerjee2021} to obtain the various couplings defined in Eq.~(\ref{eq.5}) as
\begin{subequations}\label{Eq:SecondOrderParameter}
\begin{align}
\label{eq.6.1}
J & = \frac{4}{27} \sum_{m=-\infty}^{\infty} \mathcal{J}_{m}^2 (\zeta) \Bigg[ \frac{6t_1 (t_1+2t_3)}{E_P + m\omega} + \frac{2(t_1-t_3)^2 }{E_D + m\omega}  + \frac{(2t_1+ t_3)^2 }{E_S + m\omega} \Bigg], \\
\label{eq.6.2}
K & = \frac{8}{9} \sum_{m=-\infty}^{\infty} \mathcal{J}_{m}^2 (\zeta) \frac{J_{\mathrm{H}}[(t_1-t_3)^2 -3t_2^2]}{(E_P + m\omega)(E_D + m\omega)}, \\
\label{eq.6.3}
\Gamma & = \frac{16}{9} \sum_{m=-\infty}^{\infty} \mathcal{J}_{m}^2 (\zeta) \frac{J_{\mathrm{H}}t_2 (t_1-t_3)}{(E_P + m\omega)(E_D + m\omega)} ,
\end{align}
\end{subequations}
where $\mathcal{J}_m(\zeta)$ is the Bessel function of the first kind of order $m$, $\zeta = R_{ij} E_0/\omega$ is the dimensionless amplitude of the applied laser and $R_{ij}$ is the bond length between the two TM sites. Notice that the expressions for $K$ and $\Gamma$ have nodes at the identical values of the drive strength $\zeta$ dictated by the combination of the Bessel functions and the denominator only and also are found to follow a constant relative strength as illustrated in Fig.~\ref{fig:JKGhTuningWLigand}. Therefore, it forbids the relative tunability of these parameters in the system. The comparison of our ED results with the analytical expressions for the couplings $J$, $K$, and $\Gamma$ is shown in Fig.~\ref{fig:EDPertComp}(a). We find a very good agreement between the two in this case.

We now focus on the final part of our work and obtain the spin-exchange model using perturbation theory~\cite{supp} and time-dependent Schrieffer-Wolff transformation~\cite{Banerjee2021}. In this case, the ligand degrees of freedom are taken into consideration and we have to rely on a third-order perturbation theory to properly capture the ligand effects. The perturbation theory is performed utilizing the processes illustrated in Fig.~\ref{fig:SchematicThirdOrder} and we obtain the couplings defined in Eq.~(\ref{eq.5}) as
\begin{subequations}
\begin{align}
\label{eq.7.1}
J & =  \frac{8t_{pd}^2}{27} \sum_{n,m = -\infty}^{\infty} \mathfrak{J}_{m,n} (\zeta)  \Bigg[ \frac{\sin[(m-n)\psi_0]}{\Delta + m\omega} \left( \frac{2t_1+t_3}{E_S+ l\omega} + \frac{t_1-t_3}{E_D+l\omega} + 3 \frac{t_1+t_3}{E_P+l\omega} \right) \Bigg],  \\
\label{eq.7.2}
K & = \frac{8 t_{pd}^2}{9} \sum_{n,m = -\infty}^{\infty} \mathfrak{J}_{m,n} (\zeta) \bigg[ \frac{J_{\mathrm{H}}}{(\Delta + m \omega)} \frac{\sin[(m-n)\psi_0] (t_1-t_3) -3\cos[(m-n)\psi_0]t_2}{(E_P+l\omega)(E_D+l\omega)}\bigg],  \\
\label{eq.7.3}
\Gamma & =  \frac{8 t_{pd}^2}{9} \sum_{n,m = -\infty}^{\infty} \mathfrak{J}_{m,n} (\zeta) \bigg[\frac{J_{\mathrm{H}}}{(\Delta + m \omega)} \frac{ \cos[(m-n)\psi_0](t_1 -t_3) + \sin[(m-n)\psi_0]t_2 }{(E_P+l\omega)(E_D+l\omega)} \bigg], \\
\label{eq.7.4}
h & = \frac{4t_{pd}^2}{9} \sum_{n,m = -\infty}^{\infty} \mathfrak{J}_{m,n} (\zeta) \frac{\sin[(m-n)\psi_0]}{\Delta + m \omega} \bigg[ \frac{t_1 - t_3}{E_P+l\omega} +  \frac{t_1 - t_3}{E_D+l\omega} \bigg],
\end{align}
\end{subequations}
where $\mathfrak{J}_{m,n} (\zeta) = \mathcal{J}_{m+n}(\zeta)\mathcal{J}_{-m}(\zeta r_{ij}/R_{ij}) \mathcal{J}_{-n}(\zeta r_{ij}/R_{ij})$, $r_{ij}$ is the bond-length between the TM and ligand sites, $l = m+n$ and $\psi_0$ is the angle between the TM-TM and TM-ligand bond. Note that previously $\psi_0$ was considered to be equal to $\pi/4$. We observe that in the static case $\zeta = 0$ the super-exchange coupling $J$ and the Zeeman magnetic field $\mathbf{h}$ vanish. This \textit{vanishing of super-exchange} coupling is consistent with the Jackeli-Khaliluin formalism~\cite{PhysRevLett.102.017205} and the absence of $\mathbf{h}$ at $\zeta = 0$ is consistent with the intact \textit{time-reversal symmetry} of the multi-orbital system. 

We further observe the distinct analytical structure of the Kitaev term $K$ and the anisotropy $\Gamma$ [Eqs.~(\ref{eq.7.2},\ref{eq.7.3})] which dictates that the zeros of these parameters occur at different drive strength $\zeta$. This is in stark contrast to the results discussed without ligand in Eqs.~(\ref{eq.6.2},\ref{eq.6.3}) and is consistent with the variation of these parameters as shown in  Fig.~\ref{fig:JKGhTuning_Ligand}. Fig.~\ref{fig:EDPertComp}(b) shows the results for the relative comparison between the perturbation [adding the contributions from Eq.~(\ref{eq.6.1}-\ref{eq.6.3}) to Eq.~(\ref{eq.7.1}-\ref{eq.7.4})] and ED calculations. In this case, we find an excellent agreement for the case of the magnetic field $\mathbf{h}$, but have a relatively poor agreement in the case of the other parameters. Note that with the inclusion of ligand, there are many more paths possible for the super-exchange in the high order processes. Therefore, this deviation can possibly be accounted for by higher-order contributions, which cannot be neglected due to larger values of $t_{pd}$, whereas our perturbation term accounts only till third order. On the other hand, the better agreement of the perturbation results and the ED calculations for the Zeeman magnetic field $\mathbf{h}$ is justified. In this case, there is only one process, in the next order perturbation, which contributes and is roughly proportional to $t^{4}_{pd}/\Delta^2U$. Within our parameter choice, this term is much smaller than the second or third-order contributions and can safely be neglected.

\section*{Discussion}

In our discussion above, we have not taken into account the role of heating due to laser irritation. When the laser frequency is close resonances of the system, electrons can be excited efficiently resulting in severe heating. Under the drive, the model Hamiltonian can be divided into different Floquet Hamiltonian sectors according to the number of the photon, $m$, the system absorbs/emits. Initially, electrons occupy the $m=0$ sector. The laser drives the electrons to other Floquet sectors, which causes heating. Meanwhile, the effective Floquet Hamiltonian at $m=0$ sector is modified through hybridization with other sectors. There are resonances within the same Floquet sector. In our system, at the single-ion level, we have crystal field splitting between $t_{2g}$ and $e_g$ orbitals, and splitting effective $J_{\mathrm{eff}}=3/2$ and $J_{\mathrm{eff}}=1/2$ manifold due to the spin-orbit coupling. The frequency of the laser should be tuned away from these resonances. In the Mott insulator, one dominant mechanism of heating is the excitation of the doublons by laser. This resonance occurs when the photon energy is close to the energy difference between different Hubbard bands. The population of the doublons increases continuously with time in the presence of laser irritation, which eventually invalidates the effective spin Hamiltonian description. It is calculated in Ref.~\cite{PhysRevB.99.205111} that the rate of doublon generation can be very low when the laser frequency is away from resonance set by the lower and upper Hubbard band. The effective spin Hamiltonian we derived here is valid in a long time window, where the effect of doublons on magnetic interactions is also negligible.  This long time window is known as the Floquet prethermal region, which can be exponentially long in time before the system enters into a featureless infinite temperature region due to heating~\cite{Weidinger2017,PhysRevResearch.1.033202,PhysRevB.99.205111,PhysRevB.97.245122,PhysRevLett.115.256803,PhysRevLett.116.120401,PhysRevB.95.014112,PhysRevX.7.011026}.

We remark that not all the magnetic interactions can be tuned independently because we only have two tuning parameters, i.e. amplitude and frequency, of the incident circularly polarized light. As demonstrated explicitly in Fig. \ref{fig:JKGhTuning_Ligand}, we can achieve the region where $K$ is dominant over all other magnetic interactions, thus realizing a favorite condition for the Kitaev QSL. Furthermore, the Kitaev interaction $K$ can be tuned to be ferromagnetic or antiferromagnetic by light in a single material. It is believed that RuCl$_3$ and irritates are described by a ferromagnetic Kiteav interaction without a light drive. The antiferromagnetic Kitaev model has attracted considerable interest recently because it may host a new gapless QSL in the intermediate magnetic field region~\cite{PhysRevB.83.245104,PhysRevB.97.241110,PhysRevB.98.014418,PhysRevB.98.060416,hickey_emergence_2019,PhysRevB.99.140413}. Both the antiferromagnetic Kitaev interaction and magnetic field can be induced by light, thus the driven system allows us to access this gapless QSL. We can also tune into a parameter region where $\Gamma$ interaction is dominant, where a new type of QSL has been suggested recently~\cite{luo_gapless_2021}.

So far, we have focused on the circularly polarized light, our results can be readily generalized to the case with a linearly polarized light. In this case, the effective Zeeman field due to the inverse Faraday effect is absent because the time-reversal symmetry is preserved. It is natural to expect one can tune the system into an anisotropic spin Hamiltonian by linearly polarized light, which can also support QSL~\cite{UKUMAR2021}.

While we are working on the current project, there appear two nice theoretical works~\cite{Sriram2021,PhysRevB.103.L100408,strobel_comparing_2021} on the tuning of magnetic interactions in Kitaev quantum magnets by light recently, which have some overlap with our current work. In Refs.~\cite{PhysRevB.103.L100408,strobel_comparing_2021}, the authors treat the ligands by an effective hopping between two transition metal sites, and then employ Floquet theory for the two site problem. In this treatment, the ratio between K and $\Gamma$ interaction is fixed and cannot be tuned by laser. The effects of time reversal symmetry breaking and hence the inverse Faraday effect is absent because the direct hopping of electrons between two sites does not produce a net hopping phase associated with the circularly polarized light.  In Ref.~\cite{Sriram2021}, the effective magnetic interactions and the Zeeman field due to the inverse Faraday effect were calculated numerically. The authors also determined the magnetic phase diagram by exact diagonalization of the effective spin Hamiltonian numerically.  In our work, we have derived analytical expressions of these magnetic interactions and effective Zeeman field, which provides new insights on how the magnetic interactions are tuned by light.

In conclusion, we show that light can tune magnetic interactions in Kitaev quantum magnets by explicitly calculating an effective spin Hamiltonian from the multi-orbit spin-orbit coupled Hubbard model in the presence of a circularly polarized light. We employ both the exact diagonalization of the Floquet Hamiltonian and analytical  perturbation theory. We demonstrate that magnetic interactions favorite for QSL can be achieved by tuning the frequency and amplitude of the light. Our work points to a promising route to stabilize quantum spin liquid by coupling quantum magnets to light.

\section*{Methods} 
\subsection*{Floquet Theory}
We use Bloch wave theory, $|\psi(t)\rangle = e^{-i\epsilon_\alpha t}|\psi_\alpha(t)\rangle$ and solve the Schrodinger equation given by $H(t) |\psi(t)\rangle =   i\frac{\partial }{\partial t} |\psi(t)\rangle $. Using the Fourier transform given by $H_l =\frac{1}{T} \int_0^T e^{il\omega t} H(t) dt$ and $|\psi_{\alpha, m}\rangle = \frac{1}{T} \int_0^T e^{im\omega t} |\psi_\alpha(t)\rangle$, we can solve the Schrodinger equation, where the solution is given by, $\big(\epsilon_\alpha + m\omega \big)|\psi_{\alpha, m} \rangle =\sum_{m'} H_{m-m'}|\psi_{\alpha, m'}\rangle$. Here, $\alpha$ are the basis states for the singly occupied, doubly occupied states on the TM and the singly occupied states on the ligand. The results for the quasi-energies converge quickly for off-resonance and in our case $|m|<8$, is sufficient to get converged results. We use $m=0$-Floquet sector to estimate the the parameters in the spin model.

\subsection*{Formalism for Perturbation results}
We evaluate the perturbation results for the system without ligands using the second order perturbation theory~\cite{PhysRevLett.112.077204,UKUMAR2021}. The effective Hamiltonian is given by
\begin{equation}\label{eq:SecondOrderPert}
\mathcal{H}^{(2)}_{\mathrm{eff}} = - \sum_{\alpha, \beta} \sum_{n\neq 0 } \langle \beta|\mathcal{T}_{\text{dd}} \frac{|n\rangle \langle n|}{E_n-E_0}\mathcal{T}_{\text{dd}}|\alpha\rangle |\beta\rangle \langle \alpha|.
\end{equation}
Here, $\alpha, \beta \in \Big\{ |+\tfrac{1}{2}, +\tfrac{1}{2}\rangle, |+\tfrac{1}{2}, -\tfrac{1}{2}\rangle, |-\tfrac{1}{2}, +\tfrac{1}{2}\rangle, |-\tfrac{1}{2}, -\tfrac{1}{2}\rangle \Big\}$.  The $t_{2g}$-orbitals have $l_\text{eff}=1$. The $J_\mathrm{eff} =1/2$ states for a site in the orbital basis are given by $|+\tfrac{1}{2}\rangle= \tfrac{1}{\sqrt{3}}\big( d_{yz, \downarrow}^\dagger+ i d_{zx, \downarrow}^\dagger + d_{xy, \uparrow}^\dagger\big) |0\rangle $, $|-\tfrac{1}{2}\rangle=  \tfrac{1}{\sqrt{3}}\big( d_{a,yz, \uparrow}^\dagger -i d_{a,zx, \uparrow}^\dagger- d_{a,xy, \downarrow}^\dagger\big) |0\rangle$. $|n\rangle$ are the intermediates states, which consists of double occupancy on a TM site, are  $(L, S) = [(0,0), (1,1), (2,0)]$, here, $L$ and $S$ are the total spin and angular momentum of the doubly occupied site. Here $\mathcal{T}_{dd}$ is the hopping between the TM sites.

For the system with ligands, the second order results are the same as Eq. \eqref{eq:SecondOrderPert}. The third order corrections are evaluated using
\begin{equation}\label{eq:ThirdOrderPert}
\mathcal{H}^{(3)}_{\mathrm{eff}} = \sum_{\alpha, \beta} \sum_{n\neq 0 } \langle \beta|\mathcal{T}_{\text{dd}}\frac{|n\rangle \langle n|}{E_n-E_0}\mathcal{T}_{\text{pd}}\frac{|L\rangle \langle L|}{E_L-E_0}\mathcal{T}_{\text{pd}}|\alpha\rangle |\beta\rangle \langle \alpha|.
\end{equation}
Here, $|L\rangle$ (${E_L}$) are the eigen-states (energies) for the occupancy on the ligand sites. $\mathcal{T}_{pd}$ is the hopping between the TM and the ligand sites. The parameters, $J, K, \Gamma$ and $h$ evaluated using this formalism are given in Eqs.~(\ref{eq.7.1}-\ref{eq.7.4}). 

\bibliography{ref}

\section*{Acknowledgements}

This work was carried out under the auspices of the US DOE NNSA under Contract No. 89233218CNA000001 through the LDRD Program. SZL was also supported by the US Department of Energy, Office of Science, Basic Energy Sciences, Materials Sciences and Engineering Division, Condensed Matter Theory Program.

\section*{Author contributions statement}
U.K. and S.Z.L. planned the project. U.K. developed the Floquet Exact Diagonalization code and performed all of the numerical simulations. U.K. together with S.B. carried out perturbation calculations. U.K., S.B. and S.Z.L. discussed the results and wrote the manuscript. 

\section*{Additional information}
\pagebreak
\addtocounter{page}{-6}




\section*{Supplementary material}

# Supplemental for "Floquet engineering of Kitaev quantum magnets"

Umesh Kumar and Saikat Banerjee
*Theoretical Division, T-4, Los Alamos National Laboratory, Los Alamos, New Mexico 87545, USA*

Shi-Zeng Lin
*Theoretical Division, T-4 and CNLS, Los Alamos National Laboratory, Los Alamos, New Mexico 87545, USA*
(Dated: November 1, 2021)

We present the details of the mapping of the multi-orbital Hubbard-Kanamori Hamiltonian to the effective spin model. The multi-orbital Hamiltonian is given by $\mathcal{H} = \mathcal{H}_K(t) + \sum_{i=1,2} \mathcal{H}_i^C$, where

$$\mathcal{H}_K(t) = \sum_{\langle \alpha,\beta \rangle \sigma} e^{i\delta F(t)} t_{\alpha\beta} d^\dagger_{1\alpha\sigma} d_{2\beta\sigma} + t_{pd} \sum_\sigma \left( e^{i\delta F_x(t)} d^\dagger_{1zx\sigma} p_{1\sigma} - e^{iF_y(t)} p^\dagger_{1\sigma} d_{2yz\sigma} + e^{i\delta F_y(t)} d^\dagger_{1yz} p_{2\sigma} - e^{iF_x(t)} p^\dagger_{2\sigma} d_{2zx\sigma} \right) + \text{h.c.}, \quad (1)$$

where $t_{pd}$ is the hopping amplitude between the $p$- and $d$-orbitals and the hopping matrix $t_{\alpha\beta}$ between the three $d$-orbitals $t_{\alpha\beta}$ is obtained through Slater-Koster [S] interatomic matrix elements as

$$\mathcal{T}_{dd} = \sum_{\alpha\beta\sigma} t_{\alpha\beta} d^\dagger_{1\alpha\sigma} d_{2\beta\sigma} = \sum_\sigma \begin{bmatrix} d^\dagger_{1yz\sigma} & d^\dagger_{1zx\sigma} & d^\dagger_{1xy\sigma} \end{bmatrix} \begin{bmatrix} t_3 & t_2 & 0 \\ t_2 & t_3 & 0 \\ 0 & 0 & t_1 \end{bmatrix} \begin{bmatrix} d_{2yz\sigma} \\ d_{2zx\sigma} \\ d_{2xy\sigma} \end{bmatrix}. \quad (2)$$

This multi-orbital Hamiltonian is mapped to the spin Hamiltonian which is given by

$$\mathcal{H}_{\text{eff}} = \begin{bmatrix} S_1^x & S_1^y & S_1^z \end{bmatrix} \begin{bmatrix} J & \Gamma & \Gamma' \\ \Gamma & J & \Gamma' \\ \Gamma' & \Gamma' & J+K \end{bmatrix} \begin{bmatrix} S_2^x \\ S_2^y \\ S_2^z \end{bmatrix} + \mathbf{h} \cdot (\mathbf{S}_1 + \mathbf{S}_2) + \mathcal{E} n_1 n_2. \quad (3)$$

where we have included a Zeeman coupling term to account for the inverse Faraday effect due to the circularly polarized light. [B] This spin Hamiltonian can again be written in the ladder operators and is given by [the magnetic field is finite only along the $z$-axis for the bond considered here],

$$\mathcal{H}_{\text{eff}} = \frac{J}{2}(S_1^+ S_2^- + S_1^- S_2^+) + (K+J) S_1^z S_2^z + \frac{\Gamma}{2i}(S_1^+ S_2^+ - S_1^- S_2^-) + \mathcal{E} n_1 n_2 + h\left(S_1^z + S_2^z\right). \quad (4)$$

We start with the case for the system without ligand followed by the detailed presentation for the case with ligand degrees of freedom.

## I. MAPPING HUBBARD MODEL TO SPIN HAMILTONIANS: TWO TM SITES

We start with a two-site problem as present in Ref. [R]. The Kanamori Hamiltonian in the limit $U, J_H \ll \lambda \ll t$ for the two site atoms is given by

$$\mathcal{H}_1^C + \mathcal{H}_2^C = \frac{U - 3J_H}{2}[(N_1 - 5)^2 + (N_2 - 5)^2] - 2J_H(S_1^2 + S_2^2) - \frac{J_H}{2}(L_1^2 + L_2^2). \quad (5)$$

The effective Hamiltonian in the second order perturbation theory (assuming $t_1, t_2, t_3 \ll U, J_H$) is given by

$$\mathcal{H}_{\text{eff}}^{(2)} = -\sum_{\alpha,\beta} \sum_{n \neq 0} \langle \beta | \mathcal{T}_{dd} \frac{|n\rangle\langle n|}{E_n - E_0} \mathcal{T}_{dd} |\alpha\rangle |\beta\rangle\langle \alpha|. \quad (6)$$

Note that we have the initial and final states denoted by the *greek* letters $|\alpha\rangle, |\beta\rangle$ and the doublet intermediate states denoted by the *latin* letters $|n\rangle$. The explicit forms are given as (i) The ground state with $N_1 = N_2 = 1$ has $L = 1$ and $S = 1/2$ and consists of four configurations, given by; $\alpha, \beta \in \left\{ |+\frac{1}{2}, +\frac{1}{2}\rangle, |+\frac{1}{2}, -\frac{1}{2}\rangle, |-\frac{1}{2}, +\frac{1}{2}\rangle, |-\frac{1}{2}, -\frac{1}{2}\rangle \right\}$. These states are constructed from $j_{\text{eff}} = 1/2$ states for site-$i$ in the orbital basis, which are given by

$$\left|+\frac{1}{2}\right\rangle_i = \frac{1}{\sqrt{3}}(d^\dagger_{i,yz,\downarrow} + i d^\dagger_{i,zx,\downarrow} + d^\dagger_{i,xy,\uparrow})|0\rangle_i = \left(\sqrt{\frac{2}{3}} d^\dagger_{i,+,\downarrow} - i\sqrt{\frac{1}{3}} d^\dagger_{i,0,\uparrow}\right)|0\rangle_i, \quad (7a)$$

$$\left|-\frac{1}{2}\right\rangle_i = \frac{1}{\sqrt{3}}(d^\dagger_{i,yz,\uparrow} - i d^\dagger_{i,zx,\uparrow} - d^\dagger_{i,xy,\downarrow})|0\rangle_i = \left(\sqrt{\frac{2}{3}} d^\dagger_{i,-,\uparrow} + i\sqrt{\frac{1}{3}} d^\dagger_{i,0,\downarrow}\right)|0\rangle_i, \quad (7b)$$

whereas, (ii) the intermediate states ($|n\rangle$) have $N_1 = 2$ and $N_2 = 0$ on the two TM sites, where the site for $N_1 = 2$ allows for $L = \{0, 2\}$, $S = 0$ and $L = 1$, $S = 1$ and consists of 15 configurations. The two spins in the intermediate state form singlet and triplet configurations given by

$$S = 0 : |s\rangle = \frac{1}{\sqrt{2}}(|\uparrow\downarrow\rangle - |\downarrow\uparrow\rangle), \tag{8a}$$

$$S = 1 : |t_-\rangle = |\downarrow\downarrow\rangle, \ |t_0\rangle = \frac{1}{\sqrt{2}}(|\uparrow\downarrow\rangle + |\downarrow\uparrow\rangle), \ |t_+\rangle = |\uparrow\uparrow\rangle. \tag{8b}$$

Consequently, the intermediate states in the basis $|N, L, S; L_z, S_z\rangle$ and $|^{2S+1}L, L_z, S_z\rangle$ are then given by

$$|2, 0, 0; 0, 0\rangle = \frac{1}{\sqrt{3}}(|+1, -1\rangle - |0, 0\rangle + |-1, +1\rangle)|s\rangle = |^1S, 0, 0\rangle, \tag{9a}$$

$$|2, 2, 0; +2, 0\rangle = (|+1, +1\rangle)|s\rangle = |^1D, +2, 0\rangle, \tag{9b}$$

$$|2, 2, 0; -2, 0\rangle = (|-1, -1\rangle)|s\rangle = |^1D, -2, 0\rangle, \tag{9c}$$

$$|2, 2, 0; +1, 0\rangle = \frac{1}{\sqrt{2}}(|+1, 0\rangle + |0, +1\rangle)|s\rangle = |^1D, +1, 0\rangle, \tag{9d}$$

$$|2, 2, 0; -1, 0\rangle = \frac{1}{\sqrt{2}}(|-1, 0\rangle + |0, -1\rangle)|s\rangle = |^1D, -1, 0\rangle, \tag{9e}$$

$$|2, 2, 0; 0, 0\rangle = \frac{1}{\sqrt{6}}(|+1, -1\rangle + 2|0, 0\rangle + |-1, +1\rangle)|s\rangle = |^1D, 0, 0\rangle, \tag{9f}$$

$$|2, 1, 1; +1, m_S\rangle = \frac{1}{\sqrt{2}}(|+1, 0\rangle - |0, +1\rangle)|t_m\rangle = |^3P, +1, S_z\rangle, \tag{9g}$$

$$|2, 1, 1; -1, m_S\rangle = \frac{1}{\sqrt{2}}(|0, -1\rangle - |-1, 0\rangle)|t_m\rangle = |^3P, -1, S_z\rangle, \tag{9h}$$

$$|2, 1, 1; 0, m_S\rangle = \frac{1}{\sqrt{2}}(|+1, -1\rangle - |-1, +1\rangle)|t_m\rangle = |^3P, 0, S_z\rangle. \tag{9i}$$

In the above equation, $m \in \{0, \pm\}$ leads to a total of 15 intermediate states. The Slater determinant of the two occupied orbitals needs to be written in a superposition of the intermediate states given above. For the case of $S_z = 0$ and $L_z \neq 2$, the Slater determinant are given by

$$d^\dagger_{+,\uparrow} d^\dagger_{0,\downarrow} : \quad |+\uparrow, 0\downarrow\rangle - |0\downarrow, +\uparrow\rangle = |^3P, +1, 0\rangle + |^1D, +1, 0\rangle, \tag{10a}$$

$$d^\dagger_{+,\downarrow} d^\dagger_{0,\uparrow} : \quad |+\downarrow, 0\uparrow\rangle - |0\uparrow, +\downarrow\rangle = |^3P, +1, 0\rangle - |^1D, +1, 0\rangle. \tag{10b}$$

Similarly, one can express the other doublets in the $|^{2S+1}L, L_z, S_z\rangle$. Finally, using Eq. (5), we write the energies of the intermediate states as

$$|L = 0, S = 0\rangle : \quad E_S = E_2(^1S) + E_2(^1S) - 2E_1(^2P) = U + 2J_H, \tag{11a}$$

$$|L = 1, S = 1\rangle : \quad E_P = E_2(^3P) + E_2(^1S) - 2E_1(^2P) = U - 3J_H, \tag{11b}$$

$$|L = 2, S = 0\rangle : \quad E_D = E_2(^1D) + E_2(^1S) - 2E_1(^2P) = U - J_H. \tag{11c}$$

### A. Overlaps of single with Double occupancy on the TM sites

The hopping operators allow for an overlap between the single and double occupied TM sites. Here we focus on the case without ligand. This can be understood as if the hopping through the ligands has been integrated out by using $t_{pd} = 0$. The hopping can be written as $\mathcal{T}_{dd} = \mathcal{T}_1 + \mathcal{T}_2$, where $\mathcal{T}_1$ represents hopping between the same kind orbitals and is given by

$$\mathcal{T}_1 = \sum_\sigma t_1(d^\dagger_{1,zx,\sigma} d_{2,zx,\sigma} + d^\dagger_{1,yz,\sigma} d_{2,yz,\sigma}) + t_3 \sum_\sigma d^\dagger_{1,xy,\sigma} d_{2,xy,\sigma} = \sum_\sigma t_1(d^\dagger_{1,+,\sigma} d_{2,+,\sigma} + d^\dagger_{1,-,\sigma} d_{2,-,\sigma}) + t_3 \sum_\sigma d^\dagger_{1,0,\sigma} d_{2,0,\sigma}, \tag{12}$$

and $\mathcal{T}_2$ gives the inter-orbital hopping and can be written as

$$\mathcal{T}_2 = t_2 \sum_\sigma (d^\dagger_{1,zx,\sigma} d_{2,yz,\sigma} + d^\dagger_{1,yz,\sigma} d_{2,zx,\sigma}) = -it_2 \sum_\sigma (d^\dagger_{1,+,\sigma} d_{2,-,\sigma} - d^\dagger_{1,-,\sigma} d_{2,+,\sigma}). \tag{13}$$

Note that we expressed the hopping operator in the $L_z$ and $S_z$ basis. This is an easier basis to work with since we want to express the action of these operators in terms of the intermediate states. Using the definition of the states in Eq. (7a-7b), we evaluate the action of hopping operators $\mathcal{T}_1$ and $\mathcal{T}_2$ on the singly occupied basis. The action of hopping creates double occupancy on one of the two sites which is expressed as a Slater determinant of the two orbitals which is then expressed in the $|^{2S+1}L, L_z, S_z\rangle$ basis. We here prepare the intermediate state created by the action of the hopping operator $\mathcal{T}_1$ as

$$\mathcal{T}_1 \left| +\frac{1}{2} \right\rangle_1 \left| +\frac{1}{2} \right\rangle_2 = i \frac{(t_1 - t_3)}{3} \left( |^3P, +1, 0\rangle - |^1D, +1, 0\rangle \right) |0\rangle_2, \tag{14a}$$

$$\mathcal{T}_1 \left| -\frac{1}{2} \right\rangle_1 \left| -\frac{1}{2} \right\rangle_2 = i \frac{(t_1 - t_3)}{3} \left( |^3P, -1, 0\rangle_1 - \left( |^1D, -1, 0\rangle_1 \right) \right) |0\rangle_2, \tag{14b}$$

$$\mathcal{T}_1 \left| +\frac{1}{2} \right\rangle_1 \left| -\frac{1}{2} \right\rangle_2 = -\frac{t_1}{3} \left[ -\sqrt{2} \left( |^3P, 0, 0\rangle - \frac{1}{\sqrt{3}} |^1D, 0, 0\rangle - \frac{\sqrt{2}}{\sqrt{3}} |^1S, 0, 0\rangle \right) + i\sqrt{2} |^3P, -1, +1\rangle_1 \right]$$
$$- \frac{t_3}{3} \left[ -\sqrt{2} i |^3P, +1, -1\rangle_1 - \left( \frac{\sqrt{2}}{\sqrt{3}} |^1D, 0, 0\rangle - \frac{1}{\sqrt{3}} |^1S, 0, 0\rangle \right) \right], \tag{14c}$$

$$\mathcal{T}_1 \left| -\frac{1}{2} \right\rangle_1 \left| +\frac{1}{2} \right\rangle_2 = -\frac{t_1}{3} \left[ \sqrt{2} \left( |^3P, 0, 0\rangle - \frac{1}{\sqrt{3}} |^1D, 0, 0\rangle - \frac{\sqrt{2}}{\sqrt{3}} |^1S, 0, 0\rangle \right) + i\sqrt{2} |^3P, +1, -1\rangle_1 \right]$$
$$- \frac{t_3}{3} \left[ -\sqrt{2} i |^3P, -1, +1\rangle_1 + \left( \frac{\sqrt{2}}{\sqrt{3}} |^1D, 0, 0\rangle - \frac{1}{\sqrt{3}} |^1S, 0, 0\rangle \right) \right], \tag{14d}$$

while the hopping operator $\mathcal{T}_2$ generates states as

$$\mathcal{T}_2 \left| +\frac{1}{2} \right\rangle_1 \left| +\frac{1}{2} \right\rangle_2 = -\frac{t_2}{3} \left( -2i |^3P, 0, -1\rangle - |^3P, -1, 0\rangle - |^1D, -1, 0\rangle \right) |0\rangle_2, \tag{15a}$$

$$\mathcal{T}_2 \left| -\frac{1}{2} \right\rangle_1 \left| -\frac{1}{2} \right\rangle_2 = -\frac{t_2}{3} \left( -2i |^3P, 0, +1\rangle + |^3P, +1, 0\rangle + |^1D, +1, 0\rangle \right) |0\rangle_2, \tag{15b}$$

$$\mathcal{T}_2 \left| +\frac{1}{2} \right\rangle_1 \left| -\frac{1}{2} \right\rangle_2 = -\frac{t_2}{3} \left( -2i |^1D, +2, 0\rangle_1 - \sqrt{2} |^3P, +1, +1\rangle_1 \right) |0\rangle_2, \tag{15c}$$

$$\mathcal{T}_2 \left| -\frac{1}{2} \right\rangle_1 \left| +\frac{1}{2} \right\rangle_2 = -\frac{t_2}{3} \left( -2i |^1D, -2, 0\rangle_1 + \sqrt{2} |^3P, -1, -1\rangle_1 \right) |0\rangle_2. \tag{15d}$$

Plugging back the Eqs. (14a-14d) and Eqs. (15a-15d) in Eq. (6), we evaluate the expectation values of $\mathcal{H}_{\text{eff}}^{(2)}$ within the four singly occupied states *i.e.* $\langle \beta | \mathcal{H}_{\text{eff}}^{(2)} | \alpha \rangle$. Comparing the later with matrix elements $\langle \beta | \mathcal{H}_{\text{eff}} | \alpha \rangle$ where $\mathcal{H}_{\text{eff}}$ is defined in Eq. (4) we estimate the various spin exchange parameters as defined before. Note that as we consider only hopping from *site*-2 to *site*-1 through $\mathcal{T}_{dd}$ in Eq. (14a-15d), an additional factor of two is multiplied to the matrix elements. Consequently, the exchange parameters are obtained as following [*in absence of the external circularly polarized light*]

$$J = \frac{4}{27} \left( \frac{6 t_1 (t_1 + 2 t_3)}{E_P} + \frac{2(t_1 - t_3)^2}{E_D} + \frac{(2 t_1 + t_3)^2}{E_S} \right), \tag{16a}$$

$$K = \frac{8 J_H}{9 E_P E_D} \left[ (t_1 - t_3)^2 - 3 t_2^2 \right], \tag{16b}$$

$$\Gamma = \frac{16 J_H}{9 E_P E_D} t_2 (t_1 - t_3), \tag{16c}$$

$$\mathcal{E} = -\frac{1}{9} \left( \frac{5 t_1^2 + 3 t_3^2 - 2 t_1 t_3 + 7 t_2^2}{E_P} + \frac{5(t_1 - t_3)^2 + 15 t_2^2}{3 E_D} + \frac{(2 t_1 + t_3)^2}{3 E_S} \right). \tag{16d}$$

The above expressions are consistent with the results provided in Eq. (8) of the Ref. [R]. We now move on to the detailed description of the analysis in the presence of circularly polarized light. The corresponding Floquet formalism is given in the following section.

### B. Floquet formalism

The periodic time dependence Hamiltonian is restricted only to the kinetic part of the total Hamiltonian in our multi-orbital system as shown by the Peierls substitution in Eq. (1) [note that the ligand degrees of freedom is ignored, so $t_{pd} = 0$ in Eq. (1)]. We consider a circularly polarized light [*see main text*] as $\mathbf{F}(t) = E_0 (\hat{\mathbf{x}} \sin \omega t - \hat{\mathbf{y}} \cos \omega t)/\omega$ and two TM sites located at $\mathbf{R}_1 = (0, 0)$

and $\mathbf{R}_2 = (\frac{1}{\sqrt{2}}, \frac{1}{\sqrt{2}})$. Consequently, the phase is given by $\delta F(t) = \zeta \sin(\omega t - \frac{\pi}{4})$, where $\zeta = R_{12} E_0/\omega$ and $R_{12} = |\mathbf{R}_1 - \mathbf{R}_2|$ is the nearest-neighbour distance between the two TM sites.

Next we consider the Fourier transform of the time-dependent Hamiltonian as

$$H_l = \int_0^T dt \, e^{-il\omega t} \mathcal{H}_K(t) = \mathcal{J}_l(\zeta) \, e^{-il\pi/4} \sum_{\langle \alpha, \beta \rangle \sigma} t_{\alpha\beta} d^\dagger_{1\alpha\sigma} d_{2\beta\sigma} \tag{17}$$

where $\mathcal{J}_l(\zeta)$ is the Bessel function of the first kind. Finally, we evaluate the effective Hamiltonian using the Floquet formalism given by, $\mathcal{H}^{(2)}_{\text{eff}} = P_0 \sum_{l=1}^{\infty} \frac{1}{l\omega} [H_l, H_{-l}] P_0$ where $P_0$ is the projector for the singly occupied states [K, L, B]. Note that the overall hopping phase due to the circularly polarized light dropouts in the effective Hamiltonian as the particle would hop to and from along the same bond. The spin-exchange parameters are eventually evaluated in the second-order approximation as

$$J = \frac{4}{27} \sum_{m=-\infty}^{\infty} \mathcal{J}_m^2(\zeta) \left( \frac{6t_1(t_1 + 2t_3)}{E_P + m\omega} + \frac{2(t_1 - t_3)^2}{E_D + m\omega} + \frac{(2t_1 + t_3)^2}{E_S + m\omega} \right), \tag{18a}$$

$$K = \frac{8}{9} \sum_{m=-\infty}^{\infty} \mathcal{J}_m^2(\zeta) \frac{J_H[(t_1 - t_3)^2 - 3t_2^2]}{(E_P + m\omega)(E_D + m\omega)}, \tag{18b}$$

$$\Gamma = \frac{16}{9} \sum_{m=-\infty}^{\infty} \mathcal{J}_m^2(\zeta) \frac{J_H t_2(t_1 - t_3)}{(E_P + m\omega)(E_D + m\omega)}. \tag{18c}$$

Next, we show the details of the ED calculations and elaborate on the analysis to compare the estimates of the parameters as obtained here.

**Exact Diagonalization:** Using the Floquet/Bloch theorem, the Schrödinger equation can be written as $(\epsilon_\alpha + m\omega)|\psi_{\alpha,m}\rangle = \sum_{m'} H_{m-m'}|\psi_{\alpha,m'}\rangle$ [M, E] with $\alpha \in |\pm \frac{1}{2}, \pm \frac{1}{2}\rangle$ (*single occupancy* on each TM site: 4 configurations) and $|^{2S+1}L, L_z, S_z\rangle$ (*double occupancy* on one of the TM site: 15 configurations). The quasienergy operator can be written as

$$Q_{nm}|\psi_{\alpha,m}\rangle = \begin{pmatrix} \ddots & & & & \\ & H_0 - \omega & H_{-1} & & \\ & H_1 & H_0 & H_{-1} & \\ & & H_1 & H_0 + \omega & \\ & & & & \ddots \end{pmatrix} \begin{pmatrix} \vdots \\ |\psi_{\alpha,-1}\rangle \\ |\psi_{\alpha,0}\rangle \\ |\psi_{\alpha,1}\rangle \\ \vdots \end{pmatrix}. \tag{19}$$

The full Hamiltonian ($H$) in the eigen-state basis can be written as $H = \sum_\mathfrak{n} E_\mathfrak{n} |\phi_\mathfrak{n}\rangle\langle\phi_\mathfrak{n}|$, where $|\phi_\mathfrak{n}\rangle = \sum_{\alpha,m} a^\mathfrak{n}_{\alpha,m} |\psi_{\alpha,m}\rangle$ is the eigenvector evaluated by diagonalizing the above Hamiltonian. Notice the dimension of the matrix ($H$) is $(2m_{\text{cut}} + 1) \times (4 + 15)$, where Floquet index, $m \in [-m_{\text{cut}}, m_{\text{cut}}]$ with a cutoff $m_{\text{cut}}$. To restrict the basis to $m = 0$ and singly occupied basis, one can use a projector $P_{s,0} = P_{\alpha \in s} P_{m=0}$. We then have the projected Hamiltonian ($H_{PG}$) given by

$$H_{PG} = P_{s,0} H P_{s,0} = P_{s,0} E_\mathfrak{n} (a^\mathfrak{n}_{\alpha,m})^* a^\mathfrak{n}_{\alpha',m'} |\psi_{\alpha',m'}\rangle\langle\psi_{\alpha,m}| P_{s,0} = E_\mathfrak{n} (a^\mathfrak{n}_{s,0})^* a^\mathfrak{n}_{s',0} |\psi_{s',0}\rangle\langle\psi_{s,0}|. \tag{20}$$

We know that $\mathfrak{n}$ spans over all the Floquet space where the eigen-energies are separated by $\omega$. We have restricted ourselves to the manifold corresponding to ($m = 0$)-sector. In this too, we restrict to the lowest four eigen-energies ($\alpha \in s$). The parameters say $J$ is given by

$$J = 2\langle -\frac{1}{2}, +\frac{1}{2} | H_{PG} | +\frac{1}{2}, -\frac{1}{2}\rangle = 2 \sum_{\mathfrak{n} \in \{s,0\}} E_\mathfrak{n} (a^\mathfrak{n}_{+-,0})^* a^\mathfrak{n}_{-+,0}. \tag{21}$$

Similarly, other parameters can be estimated using Eq. (4). This formalism in the limit, $t_1 = t_3$ and $J_H = 0$, recovers the single-orbital Hubbard model result reported in Ref. [M].

### C. Multiorbital system with ligand

Now we consider the ligand degree of freedom. In this case, we need to go to the third-order process. The third-order corrections are evaluated using

$$\mathcal{H}^{(3)}_{\text{eff}} = \sum_{\alpha,\beta} \sum_{n \neq 0} \langle \beta | \mathcal{T}_{\text{dd}} \frac{|n\rangle\langle n|}{E_n - E_0} \mathcal{T}_{\text{pd}} \frac{|L\rangle\langle L|}{E_L - E_0} \mathcal{T}_{\text{pd}} |\alpha\rangle |\beta\rangle\langle\alpha|, \tag{22}$$

5where $|L\rangle$ ($E_L$) are the eigen-states(energies) for the occupancy on the ligand sites. There possible eight configurations, with each ligand allowing four possible configurations; $\{|\pm\frac{1}{2}\rangle_1|\uparrow\rangle_L, |\pm\frac{1}{2}\rangle_1|\downarrow\rangle_L\}$ and the energy for each of these is given by $\Delta$. $\mathcal{T}_{pd}$ is the hopping between the TM and the ligand sites. We now consider the full time-dependent Hamiltonian as defined in Eq. (1) with both the TM and ligand hopping terms along with their Peierls phases for the externally applied circularly polarized light as before. The positions of the TM sites are chosen as $\mathbf{R}_1 = (0,0)$, $\mathbf{R}_2 = (\frac{1}{\sqrt{2}}, \frac{1}{\sqrt{2}})$ whereas the ligand sites are chosen as $\mathbf{r}_1 = (\frac{1}{\sqrt{2}}, 0)$, and $\mathbf{r}_2 = (0, \frac{1}{\sqrt{2}})$ as illustrated in Fig. 1 of the main text. Consequently, the Peierls phases are given by $\delta F(t) = \zeta \sin(\omega t - \frac{\pi}{4})$, $\delta F_x(t) = \frac{\zeta}{\sqrt{2}} \sin \omega t$, $\delta F_y(t) = -\frac{\zeta}{\sqrt{2}} \cos \omega t$, respectively. Taking the Fourier transform, we have

$$H_l = \frac{1}{T} \int_0^T dt\, \mathcal{H}_K(t) e^{-il\omega t}$$
$$= \sum_{\alpha,\beta} t_{\alpha\beta} e^{-il\pi/4} \mathcal{J}_l(\zeta) d^\dagger_{1,\alpha} d_{2,\beta} + \sum_{\alpha,\beta} t_{\alpha\beta} e^{-il\pi/4} \mathcal{J}_{-l}(\zeta) d^\dagger_{2,\alpha} d_{1,\beta}$$
$$+ \sum_\alpha \left( t_{pd} \mathcal{J}_l\left(\frac{\zeta}{\sqrt{2}}\right) d^\dagger_{1,\alpha} p_1 + e^{il3\pi/2} t_{pd} \mathcal{J}_l\left(\frac{\zeta}{\sqrt{2}}\right) d^\dagger_{1,\alpha} p_2 - e^{il3\pi/2} t_{pd} \mathcal{J}_l\left(\frac{\zeta}{\sqrt{2}}\right) p^\dagger_1 d_{2,\alpha} - t_{pd} \mathcal{J}_l\left(\frac{\zeta}{\sqrt{2}}\right) p^\dagger_2 d_{2,\alpha} \right) + (1 \Leftrightarrow 2), \quad (23)$$

One can calculate the action of hopping through ligands on the pseudospin-1/2 states for the static case. Here $p_{1,\sigma}$ and $p_{2,\sigma}$ are the edge-shared ligand $p_z$ orbitals at $\mathbf{r}_1$ and $\mathbf{r}_2$ respectively. Now we need to act the hopping operators to create doubly occupied states at site-1 and evaluate the overlaps, $\langle n|\mathcal{T}_{pd}|L\rangle\langle L|\mathcal{T}_{pd}|\alpha\rangle$. We use the relation, $d_{i,yz} = \frac{1}{\sqrt{2}}(d_{i,+} + d_{i,-})$ and $d_{i,zx} = -\frac{1}{i\sqrt{2}}(d_{i,+} - d_{i,-})$ and write the states obtained by the action of the hopping operators as

$$d^\dagger_{1,zx,\sigma} p_{1,\sigma} p^\dagger_{1,\sigma} d_{2,yz,\sigma} |+\frac{1}{2},+\frac{1}{2}\rangle = -\frac{1}{6}\left(-2i|^3P,0,-1\rangle_1 - |^3P,-1,0\rangle_1 - |^1D,-1,0\rangle_1 - |^3P,+1,0\rangle_1 + |^1D,+1,0\rangle_1\right)|0\rangle_2, \quad (24a)$$

$$d^\dagger_{1,zx,\sigma} p_{1,\sigma} p^\dagger_{1,\sigma} d_{2,yz,\sigma} |+\frac{1}{2},-\frac{1}{2}\rangle = -\frac{1}{6}\left[-2i|^1D,+2,0\rangle_1 - \sqrt{2}|^3P,+1,+1\rangle_1 - \sqrt{2}i\left(|^3P,0,0\rangle_1 - \frac{1}{\sqrt{3}}|^1D,0,0\rangle_1 - \frac{\sqrt{2}}{\sqrt{3}}|^1S,0,0\rangle_1\right)\right.$$
$$\left. - \sqrt{2}|^3P,-1,+1\rangle_1\right]|0\rangle_2, \quad (24b)$$

$$d^\dagger_{1,zx,\sigma} p_{1,\sigma} p^\dagger_{1,\sigma} d_{2,yz,\sigma} |-\frac{1}{2},+\frac{1}{2}\rangle = -\frac{1}{6}\left[-2i|^1D,-2,0\rangle_1 + \sqrt{2}|^3P,-1,-1\rangle_1 - \sqrt{2}i\left(|^3P,0,0\rangle_1 - \frac{1}{\sqrt{3}}|^1D,0,0\rangle_1 - \frac{\sqrt{2}}{\sqrt{3}}|^1S,0,0\rangle_1\right)\right.$$
$$\left. + \sqrt{2}|^3P,+1,-1\rangle_1\right]|0\rangle_2, \quad (24c)$$

$$d^\dagger_{1,zx,\sigma} p_{1,\sigma} p^\dagger_{1,\sigma} d_{2,yz,\sigma} |-\frac{1}{2},-\frac{1}{2}\rangle = -\frac{1}{6}\left(-2i|^3P,0,+1\rangle_1 + |^3P,+1,0\rangle_1 + |^1D,+1,0\rangle_1 + |^3P,-1,0\rangle_1 - |^1D,-1,0\rangle_1\right)|0\rangle_2. \quad (24d)$$

Similarly, the states obtained by the action of hopping mediated by the second ligand are given by

$$d^\dagger_{1,yz,\sigma} p_{2,\sigma} p^\dagger_{2,\sigma} d_{2,zx,\sigma} |+\frac{1}{2},+\frac{1}{2}\rangle = -\frac{1}{6}\left(-2i|^3P,0,-1\rangle_1 - |^3P,-1,0\rangle_1 - |^1D,-1,0\rangle_1 + |^3P,+1,0\rangle_1 - |^1D,+1,0\rangle_1\right)|0\rangle_2, \quad (25a)$$

$$d^\dagger_{1,yz,\sigma} p_{2,\sigma} p^\dagger_{2,\sigma} d_{2,zx,\sigma} |+\frac{1}{2},-\frac{1}{2}\rangle = -\frac{1}{6}\left[-2i|^1D,+2,0\rangle_1 - \sqrt{2}|^3P,+1,+1\rangle_1 + \sqrt{2}i\left(|^3P,0,0\rangle_1 - \frac{1}{\sqrt{3}}|^1D,0,0\rangle_1 - \frac{\sqrt{2}}{\sqrt{3}}|^1S,0,0\rangle_1\right)\right.$$
$$\left. + \sqrt{2}|^3P,-1,+1\rangle_1\right]|0\rangle_2, \quad (25b)$$

$$d^\dagger_{1,yz,\sigma} p_{2,\sigma} p^\dagger_{2,\sigma} d_{2,zx,\sigma} |-\frac{1}{2},+\frac{1}{2}\rangle = -\frac{1}{6}\left[-2i|^1D,-2,0\rangle_1 + \sqrt{2}|^3P,-1,-1\rangle_1 + \sqrt{2}i\left(|^3P,0,0\rangle_1 - \frac{1}{\sqrt{3}}|^1D,0,0\rangle_1 - \frac{\sqrt{2}}{\sqrt{3}}|^1S,0,0\rangle_1\right)\right.$$
$$\left. - \sqrt{2}|^3P,+1,-1\rangle_1\right]|0\rangle_2, \quad (25c)$$

$$d^\dagger_{1,yz,\sigma} p_{2,\sigma} p^\dagger_{2,\sigma} d_{2,zx,\sigma} |-\frac{1}{2},-\frac{1}{2}\rangle = -\frac{1}{6}\left(-2i|^3P,0,+1\rangle_1 + |^3P,+1,0\rangle_1 + |^1D,-1,0\rangle_1 - |^3P,-1,0\rangle_1 + |^1D,-1,0\rangle_1\right)|0\rangle_2. \quad (25d)$$

Note that in the case of a path through ligand, we have an additional doublet state compared to the case of $\mathcal{T}_1$ and $\mathcal{T}_2$. We further notice that the summation of the path through $L_1$ and $L_2$ recovers $\mathcal{T}_2|\alpha/\beta\rangle$ discussed in the case without a ligand owing to the cancellation of additional pathway here in the static case. However, this will not be true in the driven case due to different phase factors in the two pathways.

Using Eq. (22) and Eq. (4), we estimate the parameters using the relations

$$\langle -\tfrac{1}{2}, +\tfrac{1}{2} | \mathcal{H}_{\text{eff}} | +\tfrac{1}{2}, -\tfrac{1}{2} \rangle = \frac{J}{2}, \tag{26a}$$

$$\langle +\tfrac{1}{2}, +\tfrac{1}{2} | \mathcal{H}_{\text{eff}} | +\tfrac{1}{2}, +\tfrac{1}{2} \rangle = \frac{J+K}{4} + h + \mathcal{E}, \tag{26b}$$

$$\langle -\tfrac{1}{2}, -\tfrac{1}{2} | \mathcal{H}_{\text{eff}} | -\tfrac{1}{2}, -\tfrac{1}{2} \rangle = \frac{J+K}{4} - h + \mathcal{E}, \tag{26c}$$

$$\langle -\tfrac{1}{2}, +\tfrac{1}{2} | \mathcal{H}_{\text{eff}} | -\tfrac{1}{2}, +\tfrac{1}{2} \rangle = -\frac{J+K}{4} + \mathcal{E}, \tag{26d}$$

$$\langle +\tfrac{1}{2}, +\tfrac{1}{2} | \mathcal{H}_{\text{eff}} | -\tfrac{1}{2}, -\tfrac{1}{2} \rangle = \frac{1}{2i}\Gamma. \tag{26e}$$

We estimate the various parameters using Eq. (26a-26e) and the various matrix overlap discussed above. The Floquet phase appears only on the spatial part and there is a remnant phase due to distinct forward and backward paths. The closed hopping path through ligands now results in a nonzero hopping phase in the presence of a circularly polarized light. The hopping phase breaks time-reversal symmetry and gives a nonzero contribution to $J$, $K$, and particularly an effective magnetic field $h$ at the third-order process. The third-order term are given by [B]

$$J = \frac{8 t_{pd}^2}{27} \sum_{n,m=-\infty}^{\infty} \mathfrak{I}_{m,n}(\zeta) \left[ \frac{\sin[(m-n)\psi_0]}{\Delta + m\omega} \left( \frac{2t_1 + t_3}{E_S + l\omega} + \frac{t_1 - t_3}{E_D + l\omega} + 3 \frac{t_1 + t_3}{E_P + l\omega} \right) \right], \tag{27a}$$

$$K = \frac{8 t_{pd}^2}{9} \sum_{n,m=-\infty}^{\infty} \mathfrak{I}_{m,n}(\zeta) \left[ \frac{J_H}{(\Delta + m\omega)} \frac{\sin[(m-n)\psi_0](t_1 - t_3) - 3\cos[(m-n)\psi_0]t_2}{(E_P + l\omega)(E_D + l\omega)} \right], \tag{27b}$$

$$\Gamma = \frac{8 t_{pd}^2}{9} \sum_{n,m=-\infty}^{\infty} \mathfrak{I}_{m,n}(\zeta) \left[ \frac{J_H}{(\Delta + m\omega)} \frac{\cos[(m-n)\psi_0](t_1 - t_3) + \sin[(m-n)\psi_0]t_2}{(E_P + l\omega)(E_D + l\omega)} \right], \tag{27c}$$

$$h = \frac{4 t_{pd}^2}{9} \sum_{n,m=-\infty}^{\infty} \mathfrak{I}_{m,n}(\zeta) \frac{\sin[(m-n)\psi_0]}{\Delta + m\omega} \left[ \frac{t_1 - t_3}{E_P + l\omega} + \frac{t_1 - t_3}{E_D + l\omega} \right], \tag{27d}$$

where $\mathfrak{I}_{m,n}(\zeta) = \mathcal{J}_{m+n}(\zeta)\mathcal{J}_{-m}(\zeta r_{ij}/R_{ij})\mathcal{J}_{-n}(\zeta r_{ij}/R_{ij})$, $r_{ij}$ is the bond-length between the TM and ligand sites, $l = m + n$ and $\psi_0$ is the angle between the TM-TM and TM-ligand bond.

**Exact Diagonalization:** We use the formalism as used in the earlier case of the system without ligand. Here we include the occupancy on the ligands in the basis set. For this case we need to use the hopping overlaps from Eqs. (24a-24d) and Eqs.(25a-25d) in addition to Eqs. (14a-14d) and Eqs. (15a-15d) and then use the same formalism that is used for the two-site case.


\end{document}